# A Shift In Artistic Practices through Artificial Intelligence

**Authors**

Kıvanç Tatar*
Chalmers University of Technology
Gothenburg, Sweden
*tatar@chalmers.se*

Petter Ericson*
Umeå University
Sweden
*pettter@cs.umu.se*

Kelsey Cotton*
Chalmers University of Technology
Gothenburg, Sweden
*kelsey@chalmers.se*

Paola Torres Núñez del Prado
Stockholm University of the Arts
Stockholm, Sweden
*paola.torres@uniarts.se*

Roser Batlle-Roca
Universitat Pompeu Fabra, Spain
*roser.batlle@upf.edu*

Beatriz Cabrero-Daniel
University of Gothenburg
Gothenburg, Sweden
*beatriz.cabrero-daniel@gu.se*

Sara Ljungblad
University of Gothenburg
Sweden
*sara.ljungblad@chalmers.se*

Georgios Diapoulis
Chalmers University of Technology
Gothenburg, Sweden
*georgios.diapoulis@chalmers.se*

Jabbar Hussain
University of Gothenburg
Sweden
*jabbar.hussain@ait.gu.se*

* These authors share equal contribution as the first authors.

**Abstract**

The explosion of content generated by artificial intelligence (AI) models has initiated a cultural shift in arts, music, and media, whereby roles are changing, values are shifting, and conventions are challenged. The vast, readily available dataset of the Internet has created an environment for AI models to be trained on any content on the Web. With AI models shared openly and used by many globally, how does this new paradigm shift challenge the status quo in artistic practices? What kind of changes will AI technology bring to music, arts, and new media?

## *Current AI Applications in Artistic Practices*

Increasing interest and accessibility in artificial intelligence (AI) technologies have moved the societal discussion of AI to the mainstream. Ease in accessing computation resources, open-access AI knowledge, open-source AI software, and vast datasets in different modalities enable artists to explore new possibilities. The recent advancements in AI architectures have been applied to all artistic domains, including visual arts, video art, music and sound art, dance and performance arts, literature, and interdisciplinary arts.

AI technologies for artistic image and video generation [1] have had to build an entirely new vocabulary for describing images or frames in video. While image generation with autonomous painting was established decades ago (such as in Harold Cohen's AARON system), it is only relatively recently that AI in the visual domain has become accessible to vast communities thanks to open-source, open-access, and browser-based tools that require minimal local computation resources, such as Stable Diffusion, and generative adversarial networks such as BigGAN and StyleGAN.

In music and sound arts, there are four main tracks of AI applications [2] and interactive music systems [3]: approaches working with symbolic music; systems working with audio signals directly; systems that map a symbolic music domain to musical audio or vice versa; and approaches that connect musical domains to other domains such as bodily movement.

In performing arts, movement generation [4] has been one AI research track. The advances in this domain have created a ripple effect in generative dance and animation: AI systems have been used as a tool for assistance, choreography support, performance collaboration, and movement generation.

Text generation and conversational bots [5] have initiated recent discussions on how omnipresent large language models (LLMs) such as ChatGPT can change all aspects of society. Additionally, LLMs have been incorporated into interdisciplinary domains and interactive artworks with chatbots, as well as tool generation for other artistic domains such as audio synthesizer code or creative coding script generation examples.

There is currently no consensus on the definition and coverage of the term artificial intelligence in the literature [6]. Even though we mainly cover machine learning (ML) approaches utilizing training datasets here, these are often embedded in complex systems with sensing and action capabilities. These approaches are alternatively referred to in the literature as multi-agent systems [7]. In the remainder of the paper, we use the term AI to include systems based on ML approaches, which may or may not have sensing and action capabilities. We call those systems applied artificial intelligence technologies---or simply AI---in the remainder of the paper.

## *Infinite Content on the Boundless Internet*

There is a long history of digital art that generates infinite content in various guises [8]. New AI approaches such as Stable Diffusion and MusicLM have received widespread interest in public discussions due to their aesthetic possibilities when combined with LLMs. These AI models are trained, fine-tuned, and parameterized through a process involving many actors, including data creators, AI developers, and model architecture researchers. Artists have been joining all processes of AI technology----including data creation, developing novel AI architectures, training specific models, and utilizing readily available models. As the technical skill barrier in using AI technology has decreased due to ease of access for known and recent models, all digital platforms have been receiving a new wave of AI-generated content.

As a result of the increased accessibility of AI tools, values in artistic production have been shifting from favoring manually made content to automatically generated content. AI systems such as Stable Diffusion have been challenging conventional values in visual arts practices by generating infinite amounts of content. In some known cases, the training datasets of widely used AI content generators were gathered by scraping content from the Internet, even when the data was held within copyright. A recent Vox short documentary mentions how the American illustrator James Gurney has become a common style prompt entry in AI image generation. The artist raises the issue of consent within infinite content generators. Gurney has stated:

> *"I think it is only fair to people looking at this work that they should know what the prompt was and also what software was used. I think the artists should be allowed to opt-in or opt-out of having their work that they worked so hard on by hand be used as a dataset for creating this other artwork."*

Gurney's comment highlights the current status quo: AI technology developers overlook the question of explicit consent from artists and content owners, and current regulations fall short of empowering artists in decisions on the use of their content. Furthermore, massive content generation in the style of a particular artist could devalue it, as in the case of Gurney, or increase it by social networking, as in the case of Holly Herndon [10], who has shared her voice as an ML model (vocal deepfake) for other artists to use in their performances. This shift in capital value on cultural practices calls for a new take on ethics for artistic practices in the era of ubiquitous AI.

## *Labor in Artistic Data*

The increasing requirement of large datasets for mass generated high-fidelity content using AI has triggered a notable shift in practice of Internet ethics of copying, appropriating, and distributing art and music. Currently, the datasets that are used to train AI models such as DALL-E, Stable Diffusion, or GitHub CoPilot have been scraping data from the Web while ignoring the licenses and intellectual property of data. This is akin to a digital enclosure of the cultural commons.

The case of Gurney is one example of how the creators of these widely used, easily accessible AI models for content generation have been gathering data for their training datasets: by overlooking the consent of data creators. This directly exploits the artist's labor in creating the original content [11]. The data that is publicly shared on the Internet is approached as free to take and is transformed into a capital commodity through the process of training ML models, exhibiting similarities to historical practices of exploitatively ignoring cultural context and consent within the structures of colonialism and its enclosures. This current status quo in data gathering can be addressed through empowering the artist by incorporating their explicit consent into current structures. The consent can be manifested by the artist, as in the case of Stability.ai, where artists can choose an option so that their work is not used in the training of Stable Diffusion models. However, it is not clear to what extent this pledge can be honored, considering the multitude of remixes and format shifts of popular artists' works on the Internet. Other licenses in that discussion, such as Creative Commons, emphasize the importance of fair use for public good [12]. While both proposals highlight valuable aspects of the issue, we envisage that a future of copyright in artistic practices will require platforms and structures promoting artistic data sovereignty.

Although existing copyright licenses are legally recognized mechanisms for protecting intellectual property, the current gatekeeping mechanisms in accessing online platforms have implicit consent structures related to artistic work, such as terms of service agreements and cookie mechanisms. Those agreements are often produced without the participation of the artists, thereby benefiting the industry that is gatekeeping access to the digital platforms. Whether the copyright permissions of artists are respected through explicit consent is an issue of traceability, accountability, and regulation. Many practitioners in artistic domains are individuals or small business owners who do not have the power to counter the impositions of big-tech such as OpenAI. We still need tools of traceability, third-party nonprofit organizations for proposing regulations, and public structures for accountability. Traceability, regulation, and accountability all require inclusive and participatory discussions of ethics for the foreseeable future of AI.

## A Shift in Artistic Practices

The emergence of accessible AI tools for artistic production caused a natural shift in artistic practices. The conception of artist as genius, where a single person produces a masterpiece, has been shifting toward communal production, where several actors are involved in the artistic production. Nonprofits such as EleutherAI, and for-profits such as Stability.ai, with involvement by other developers, have been pursuing open-sourcing the groundbreaking AI architectures, which has been benefiting artistic practices by initiating public discussions.

Even though AI technology employs the conventions of open-source and open-access tools, the question remains whether the power structures within these technologies are truly democratic. The majority of tools for AI development are provided, hosted, or maintained by a few technology companies, such as Google (TensorFlow, Collab), OpenAI (DALLE-2, GPT-3), Facebook (PyTorch), and Microsoft (GitHub). Although there is a shift toward communal production in artistic practices, with varying levels of contributions to the artwork production by different actors, these technology companies still wield decisive power and capital determining the use of AI tools for artistic practices and their computational resources. Decision-making in creation of AI tools for artistic production has yet to be democratized.

An artwork exists in close relation to its medium. Along with ubiquitous AI, the evolution of new art markets such as social media, streaming platforms, and nonfungible tokens initiated a cultural and monetary value reassignment by aggregating their status as major channels of artwork "storage" and distribution. The value of an artwork is typically affected by the ranking or curation mechanisms of the medium. Those mechanisms are directly related to the artwork's visibility and thereby its value. These platforms now have substantial influence in reshaping current social conceptions around economic and cultural capital. For example, the notion of added value through engagement appears in Herndon's AI system for voice synthesis, titled Holly+. Herndon trained the AI architecture on her own voice recordings, and Holly+ is an AI model through which users can synthesize recorded audio in the style of Herndon's voice. Herndon deploys a decentralized autonomous organization (DAO) blockchain to allow voting on the minting of artworks made using Holly+ and distributes tokens [13] to members of the DAO (and the creator of minted artworks) to share in any profits from the usage of Holly+ recorded on the relevant blockchain. Here, Herndon acknowledges the value put into Holly+ by each user when they engage with the AI system. By giving away DAO tokens, Herndon creates a platform for users to both acquire a voting stake in the Holly+ DAO and to fiscally benefit from the capital value produced by artistic work made with Holly+. This semi-decentralized governing structure for Holly+ is centered on connecting value to engagement processes: The more people who engage with the art, the more valuable the art itself is perceived to be.

## *Sustainability, Ethics, Accessibility, and Inclusion*

Discussion on societal aspects of AI is entangled with the concepts of sustainability, ethics, and values. A broader framing of these issues, focused on the evolution of culture and impact upon culture industries, has implications for the environmental and climate impact of leveraging AI technologies within artistic practices. Within discussions around sustainability, there are also broader concerns with social sustainability for artists and the sustainability of the role of art within society. A core concern in this regard is how advancements in AI technologies deployed within artistic contexts may adversely impact the cultural economy of artists in society, and further, how their role in society is affected by the wider usage of AI tools.

The appreciation of art is in the eyes of the beholder. A larger discussion in this regard is centered on the intrinsic nature of AI tool usage as originating from a desire to "be better" and thereby enable "better" artistic work. This highlights questions regarding what the nature of "better" actually is, who decides this, and if this continual pursuit of "better" is an enacted artistic Darwinism. While conclusive decisions on how power, relevancy, and what "better" means are yet to be navigated, it is plainly apparent that the positioning of interdisciplinary borders is significant in shaping value systems determining the role AI occupies within cultural industries. To this end, the rigidity (or fuzziness) of these boundaries needs to facilitate equitable exchange, promote artistic diversity, and encourage an artistic culture of mindful progress.

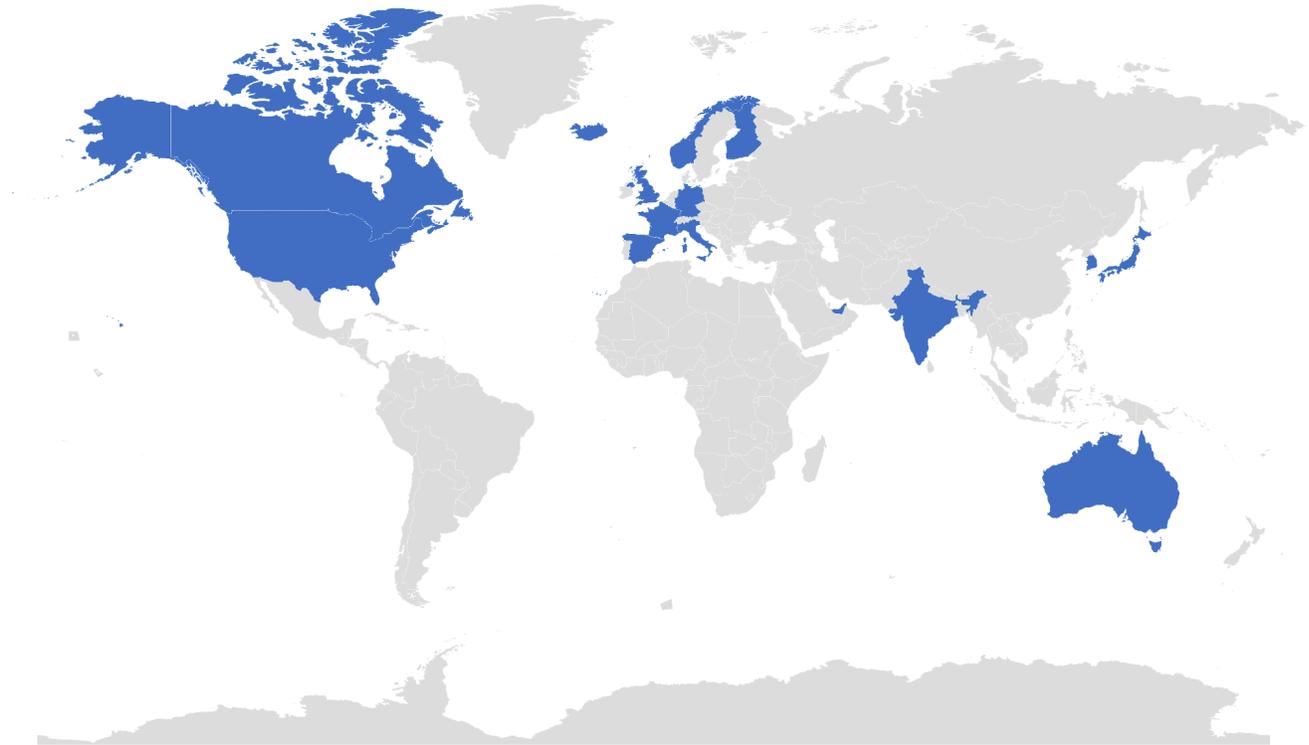

Figure 1 - Countries with at least one organization (e.g., government, university) issuing AI guidelines for responsible and ethical AI development and implementation (not necessarily laws, as of 2019): Australia, Canada, Finland, France, Germany, Iceland, India, Italy, Japan, Netherlands, Norway, Singapore, South Korea, Spain, U.A.E., U.K., and U.S.A. Populations of these countries (as of 2021): 2.8 billion, compared to the total world population of 7.9 billion. Therefore, these organizations make recommendations for 35% of the world's population [19]. Creative Commons 0 license—Public domain.

A world with more widely available AI art creation tools and globalized platforms, giving more people than ever the ability to create and share art, could lead to a greater chance to find a wider audience for previously unknown and/or marginalized creators, styles, and art forms. Currently, however, the same world and those same platforms are more often built to entrench the dominance of the hegemonic culture and existing artists. There is still more to be done to improve inclusion [14]. Geographical borders still matter in accessibility of the Internet and protection of rights in AI applications (Fig. 1). Therefore, communities must take action to address the active participation of underrepresented cultures beyond the passive appropriation of their imaginaries. Cultivating diversity requires ethical considerations in all processes of all stakeholders and actors of AI while reducing biases and analyzing stereotypes and representation of individuals and their creations.

Some ubiquitous AI models have been shown to reproduce problematic stereotypes in the data [15], which also may appear in artworks that are generated using these models. It is often hard for a single user or artist to be aware of those stereotypes in AI models, since they become apparent after many output generations on similar prompts and require statistical analysis on the generated content. In the case of tools such as DALL-E, Midjourney, and Stable Diffusion, it is worth noting the difference in the quality of the images generated when the prompted text used for generation refers to images, symbols, or art that belongs to a hegemonic culture (undoubtedly easier to access) when compared to the outputs of prompts that relate to more "obscure," underrepresented, or nonhegemonic cultural manifestations [16]. Cultural hierarchies could indeed be extrapolated from how much detail, quality, or "realism" (or lack thereof) the generated images finally contain.

## *A Better Future for Artistic Practices with Artificial Intelligence*

From data production and curation to model design, implementation, training, configuration, and final use, a human is involved at every point in AI technologies. Ethical considerations emerge in the decisions of all stakeholders and actors, in addition to the artists. The conceptualizations and values of technology actors and stakeholders leave "residues" behind that influence the artworks and the culture and society that they are situated in. AI technology, with its immense power to shift culture and society globally, goes beyond proprietary software rights of a single for-profit company. Public discussions on what a future society with AI may look like are critical.

Primarily, a fundamental change in power and the distribution of power is necessary for inclusivity in the decision-making of all AI system designs. In the case of recommendation systems, inclusive decision-making processes can have significance in the obscuring and hypervisibilizing of artworks and artists, especially those from underrepresented and marginalized communities, cultures, and regions. Additionally, a crossover between art and technology within pedagogical contexts is beneficial, allowing artists to acquire the knowledge, skills, and access to make artistic use of new AI technologies, and technologists the time, training, and opportunity to explore artistic pursuits and their requirements.

Copyright is an important aspect of artistic practice in a digital world that is further highlighted and complicated by the introduction of widely available and powerful AI technologies. Exact future steps on consent, traceability, regulations, and accountability are not clear yet. However, the discussions within, for example, various pirate parties, digital rights groups, hacktivist organizations, and open-access global initiatives of technology replications (such as GPT-NEO, GPT-J, Stable Diffusion, and DALL-E 2 PyTorch replication) are significantly more instructive than the various rearguard actions by other industry organizations. It is evident that there is an immense need for reform of the current regime of might makes right, wherein large corporations can infringe on copyright with absolute impunity, shedding light on problematic power relations. The discrepancy between corporate copyright transgressions---in the creation of enormous datasets as input to AI systems with no knowledge or consent from the copyright holders---and the automated "copyright" takedowns of private individuals' meticulously researched fair-use remix and commentary works on media platforms highlights the differences in power and punishment for private versus industrial use. Legislation, communal guidelines, and ethical dimensions of AI technologies for artistic practices are ongoing societal discussions. At the time of writing of this paper, Italy has banned the popular ChatGPT, citing security concerns for Italian citizens, in a move that may be followed by other countries, and will not allow access to this new technology until their data protection to-do list is implemented [17], whereas countries such as Japan are considering softer regulations in which content usage for AI model training is widely allowed [18]. As an alternative to draconian measures, it is long overdue, and this is the perfect moment in time to include artists and practitioners proactively in these discussions. Accommodating and fusing different voices and knowledge is a must for the reformation of equity, equality, and justice in AI technology creation. Art is for everyone, and the tools we use to make art, especially AI tools, should enable and empower just and equitable creation.

## *Acknowledgement*

The first, second, and third authors share lead authorship. This work was partially supported by the Wallenberg AI, Autonomous Systems and Software Program—Humanity and Society (WASP-HS), funded by the Marianne and Marcus Wallenberg Foundation and the Marcus and Amalia Wallenberg Foundation.

[19] Jobin, A., Ienca, M., & Vayena, E., "The global landscape of AI ethics guidelines," *Nature Machine Intelligence*, Volume 1, Issue 9, Article 9, (2019), doi: [10.1038/s42256-019-0088-2](10.1038/s42256-019-0088-2)

## *Bios*

**Kıvanç Tatar** is an musician, artist, technologist, and researcher who works in the intersection of machine learning, artificial intelligence, music, interactive arts, and design. The computational approaches developed through that interdisciplinary research have been integrated into musical and audiovisual performances, interactive artworks, and immersive environments including virtual reality. Tatar is currently an assistant professor and a WASP-HS fellow at Chalmers University of Technology, starting a new research group connecting art, music, technology, and artificial intelligence.

**Petter Ericson** is a postdoctoral fellow in the research group for Responsible AI at Umeå University in Sweden, working on graph problems and formal descriptions of structured data, with a strong interest in ethics, music, and society. His recent research interests include anti-capitalist artificial intelligence and circumventing political and structural barriers that bar AI from being used to support democratic and egalitarian values. Outside of academia, his musical interests have led him to everything from seedy late-night jam sessions at Copenhagen jazz clubs to organizing 24-hour hackathons around producing electronic music from ESA data.

**Kelsey Cotton** is a vocalist-artist-mover working with experimental music, musical artificial intelligence and human-computer interaction. Passionate about somatic interaction, the potential for intersomatic experiences between fleshy and synthetic bodies, and first-person feminist perspectives of musical AI, Cotton is pursuing doctoral studies in interactive music and AI at Chalmers University of Technology.

**Paola Torres Núñez del Prado** is pursuing doctoral studies at the Stockholm University of the Arts, focusing on researching and developing hybrid interactive textile sound interfaces that include the use of AI systems.

**Roser Batlle-Roca** is pursuing doctoral studies in transparent AI and music generation at Universitat Pompeu Fabra in Spain, in collaboration with JRC-EC and Sony.

**Beatriz Cabrero-Daniel** is a postdoctoral fellow at Gothenburg University in Sweden, currently researching trustworthy AI.

**Sara Ljungblad** is a researcher and senior lecturer at Gothenburg University and Chalmers University of Technology in Sweden, doing critical robotics, studying people's experiences and use of robotic products and autonomous systems in everyday settings in the field of human-robot interaction.

**Georgios Diapoulis** is pursuing doctoral studies in gestural interaction with generative algorithms for machine musicianship at Chalmers University of Technology and University of Gothenburg, Gothenburg, Sweden.

**Jabbar Hussain** is pursuing doctoral studies at Gothenburg University in Sweden in Informatics in the area of trustworthy/responsible AI.